\documentclass[12pt]{iopart}
\newcommand{\I}{{\rm i}}

\begin{document}

\title{Semiclassical spectral correlator in quasi one-dimensional systems}
\author{Petr Braun$^{1,2}$, Sebastian M\"uller$^3$ and Fritz Haake$^1$}
\address{$^1$ Fachbereich Physik, Universit{\"a}t Duisburg-Essen,
  47048 Duisburg, Germany}
\address{$^2$ Institute of Physics, Saint-Petersburg University,
  198504 Saint-Petersburg, Russia}
\address{$^3$ Department of Mathematics, University of Bristol, Bristol BS8 1TW, United
Kingdom}

\ead{Petr.Braun@uni-due.de}

%\maketitle

\begin{abstract}

We investigate the spectral statistics of chaotic quasi one
dimensional systems such as long wires. To do so we represent the
spectral correlation function $R(\epsilon)$ through derivatives of
a generating function and  semiclassically approximate the latter
in terms of periodic orbits. In contrast to previous work we
obtain both non-oscillatory and oscillatory contributions to the
correlation function. Both types of contributions are evaluated to
leading order in $1/\epsilon$ for systems with and without
time-reversal invariance. Our results agree with expressions from
the theory of disordered systems.

\end{abstract}
\pacs{05.45.Mt, 73.21.-b, 73.20.Fz}

\section{Introduction}

In the field of quantum chaos and disorder, the behavior of quasi
one-dimensional systems such as long wires is clearly
distinguished from the behavior of ``normal'' chaotic or
disordered systems. Most importantly, quasi one-dimensional
systems display Anderson localization \cite{Anderson}, i.e., wave
functions are localized in only part of the wire and the
conductance is suppressed. Anderson localization has important
consequences for the statistics of energy levels
\cite{AltshulerShklovskii,AndreevAltshuler}. For normal systems
the energy levels tend to repel each other; the spectral
statistics is universal and agrees with predictions made by
averaging over random-matrix caricatures of the possible
Hamiltonians, according to the so called BGS conjecture
\cite{BGS}. In contrast the spectral statistics of quasi
one-dimensional systems depends on the length (and thus the
diffusion time $T_D$); in the limit of large length the energy
levels belonging to the localized wave functions become
independent and hence show Poissonian statistics.

This difference between normal and quasi one-dimensional systems
is well understood for disordered systems. Notable approaches are
based on the DMPK equation \cite{DMPK} and on the nonlinear sigma
model, a field-theoretical technique to evaluate averages over
different realizations of the disorder potential. From the latter,
localization could be extracted in \cite{LSZ,Kamenev}. The
appropriate definition of quasi one-dimensional behavior arising
in this context is that the classical diffusion time $T_{D}$
becomes comparable to or larger than the relevant quantum time
scales, in particular the Heisenberg time $T_{H}=\frac{2\pi \hbar
}{\Delta }$ where $\Delta $ is the mean level spacing. A
random-matrix model for systems of this type was considered in
\cite{FyodorovMirlin,Bible,Efetov}.

For clean chaotic systems (e.g. wires in which the classical
motion becomes chaotic due to the shape of the boundary) the
effects of quasi one-dimensionality are less well understood, and
most of the literature is restricted to normal systems. A quantity
that has attracted a lot of attention in this context is the
spectral correlation function $R(\epsilon)$. For dynamical systems
with a level density $\rho(E)$ this correlation function is
defined by
\begin{equation}\label{defR}
R(\epsilon)=
\Delta^2\Big\langle
\rho\Big(E+\frac{\epsilon\Delta}{2\pi}\Big)
\rho\Big(E-\frac{\epsilon\Delta}{2\pi}\Big)
\Big\rangle-1\,
\end{equation}
where $\epsilon$ is a real energy offset, the brackets denote an
average over the center energy $E$ and $\Delta$ is the mean level
spacing. For simplicity we assume that $\Delta$ is brought to 1 by
appropriately scaling the energy levels. Random matrix theory
(RMT) now makes predictions for $R(\epsilon)$: For systems without
time reversal invariance an average over the Gaussian Unitary
Ensemble (GUE) of RMT gives
$-\frac{1}{2\epsilon^2}+\frac{\cos2\epsilon}{2\epsilon^2}$ while
for time reversal invariant systems an average over the Gaussian
Orthogonal Ensemble (GOE) leads to infinite power series in
$\frac{1}{\epsilon^n}$ and $\frac{\cos2\epsilon}{\epsilon^n}$. The
slow (power law) decay of oscillations leads to a singularity in
the Fourier transform of $R(\epsilon)$ (the spectral form factor)
at time $t=T_H$.

To show that individual systems are faithful to these predictions
a semiclassical approach was proposed in \cite{Berry,Argaman,SR}.
The essential idea is to express $\rho(E)$ as a sum over periodic
orbits, using Gutzwiller's formula \cite{Gutzi}, and then study
the interference between contributions of these orbits. The
leading non-oscillatory contribution arises from ``diagonal''
pairs of identical (up to time reversal) orbits \cite{Berry}. The
remaining terms were accessed only recently in
\cite{SR,Oursmalltime,Ourlargetime,KM}.

In the present paper we want to generalize these new results to quasi one-dimensional systems.
For these systems the diagonal approximation to
the small-time form factor (the Fourier image of the
non-oscillatory part of $R(\epsilon)$) was evaluated by Dittrich
\cite{Dittrich}. First results on off-diagonal contributions were
obtained by Schanz and Smilansky \cite{Schanz} for one-dimensional
quantum graphs, and by Brouwer and Altland \cite{Brouwer} who
semiclassically explained localization for quasi one-dimensional
systems modelled by
 an array of quantum dots.
In contrast, we will focus on general quasi one-dimensional systems
such as long wires. We use a periodic-orbit expansion not of the
correlation function itself, but of a generating function which
yields $R(\epsilon)$ upon taking derivatives
\cite{Ourlargetime,KM}. This enables us to determine, to  leading
order in $\frac{1}{\epsilon}$, both the non-oscillatory and the
oscillatory parts of $R(\epsilon)$. In this order we see that the
effects of quasi one-dimensionality reduce to modification of the
periodic orbit sum rule suggested in \cite{Dittrich}. For systems
without time reversal invariance it suffices to perform a diagonal
approximation on the level of the generating function. In
contrast, for time-reversal invariant systems this diagonal
approximation still captures only the non-oscillatory part; the
evaluation of the oscillatory part involves off-diagonal
contributions of pairs of non-identical but similar orbits.  Both
for systems with and without time-reversal invariance we reach
agreement with results for disordered systems by Andreev and
Altshuler \cite{AndreevAltshuler}. Our results illustrate how
semiclassical methods are useful not only for describing universal
features of ``normal'' systems but also deviations from
universality.

Higher-order corrections in $1/\epsilon$ should be similarly
accessible; their calculation needs taking into account  more
complicated groups of correlated orbits introduced in previous
work on normal systems\cite{SR,Oursmalltime}, combined with
treatment of higher order effects of quasi one-dimensionality. An
extension of this approach to Anderson localization appears within
reach.

\section{Two-point spectral correlator and the generating function}

To get started we briefly review how the correlation function
$R(\epsilon)$ can be accessed through a generating function.
Following Ref.~\cite{Ourlargetime} we write $R(\epsilon)$ as the
real part of the complex correlation function
\begin{eqnarray}\label{defC}
C(\epsilon ^{+})&=&\frac{\Delta ^{2}}{2\pi ^{2}}
\left\langle
{\rm Tr}\left( E+\frac{\epsilon ^{+}\Delta }{2\pi }-\hat{H}\right)^{-1}
{\rm Tr}\left( E-\frac{\epsilon ^{+}\Delta }{2\pi }-\hat{H}\right)^{-1}
\right\rangle
-\frac{1}{2}\,, \nonumber\\
\epsilon^\pm&=&\epsilon\pm\I\gamma\,,\label{complexcorrelator}\\
R(\epsilon)&=&\lim_{\gamma \to 0}{\rm Re}\,C(\epsilon ^{+})\nonumber
\end{eqnarray}
and determine the latter from a generating function, the
energy-averaged combination of four spectral determinants
\begin{equation}
Z\left(\epsilon _{A}^{+},\epsilon _{B}^{-},\epsilon_{C}^{+},
\epsilon_{D}^{-}\right) =
\left\langle \frac{
\det\big(E+\epsilon_{C}^{+}-\hat{H}\big)
\det\big(E+\epsilon_{D}^{-}-\hat{H}\big)}
{\det\big(E+\epsilon_{A}^{+}-\hat{H}\big)
\det\big(E+\epsilon_{B}^{-}-\hat{H}\big)}
\right\rangle
\label{defZ}
\end{equation}
as
\begin{equation}
R(\epsilon)=
\lim_{\gamma \rightarrow 0}{\rm Re}\,C\left( \varepsilon ^{+}\right)=
-\frac{1}{2}+2\,{\rm Re}\,\lim_{\gamma \rightarrow 0}
\left.
\frac{\partial ^{2}Z}{\partial\epsilon_A^{+}\partial\epsilon_{B}^{-}}
\right\vert_{\parallel
,\times }\,.  \label{CinZ}
\end{equation}
Here the subscripts $\pm$ indicate small positive or negative
imaginary parts. The symbols $\parallel ,\times $ denote two
alternative ways of identifying the energy arguments, to be
referred to as ``columnwise'' ($\parallel$) and ``crosswise''
($\times$),
\begin{eqnarray}
\parallel\; :&&
\epsilon _{A}^{+}=\epsilon _{C}^{+}=\epsilon_A^{+},\,\epsilon
_{B}^{-}=\epsilon _{D}^{-}=-\epsilon ^{+}
\hspace{3cm} {\rm columnwise}\,,  \label{para}\\
\times :&&
\epsilon _{A}^{+}=\epsilon ^{+},\,\epsilon _{B}^{-}=
-\epsilon^{+},\epsilon _{C}^{+}=-\epsilon ^{-},\epsilon _{D}^{-}=
\epsilon^{-},\gamma \rightarrow +0\quad \,\,{\rm crosswise}\,.
\label{cro}
\end{eqnarray}

Both procedures would yield the same result for the two-point
correlator if implemented rigorously. However, we shall have to
calculate $Z$ semiclassically, and that approximation entails two
different expressions, one ($\parallel$) reproducing the
non-oscillatory part and the other ($\times$) the oscillatory part of
$R(\epsilon)$. To obtain the full result both expressions have to be
added. In \cite{KM} it was shown that this addition
can be understood naturally in terms of an improved semiclassical
approximation preserving the unitarity of the time evolution (the
Riemann-Siegel lookalike formula \cite{BerryKeating}).

The semiclassical approximation for $Z$
is based on Gutzwiller's formula for the trace of the resolvent
\begin{equation}
{\rm tr}(E^+-H)^{-1}=-\frac{i\pi E^+}{\Delta}+\sum_a F_ae^{iS_a(E^+)/\hbar}
\end{equation}
The factor proportional to $E^+$ is the smooth (Weyl) part of the
level density and the sum  taken over periodic orbits with
$S_{a},T_{a},\,F_{a}$ action, period and stability coefficient of
the $a$th orbit. Integration then yields the semiclassical
approximation of the determinant
\begin{eqnarray*}
\det \left( E^{+}-\hat{H}\right)^{-1}
&=&
\exp\Big[-\int^{E^{+}}dE\,{\rm Tr}\,( E-H)^{-1}\Big]\\
&\sim &
\exp\Big(\frac{\I\pi E^{+}}{\Delta }+\sum_{a}F_{a}\,
\e^{\,\frac{\I}{\hbar }S_{a}( E^{+})}\Big)
\end{eqnarray*}
Substituting such expansions for all four determinants in $Z$ and expanding,
e.g., $S_a(E+\epsilon_A^+)\approx S_a(E)+T_a\epsilon_A^+$
we obtain
\begin{eqnarray}
Z &\approx &\e^{\frac{\I}{2}\left(\epsilon_{A}^{+}-\epsilon_{B}^{-}
-\epsilon_{C}^{+}+\epsilon_{D}^{-}\right) }\exp \Big[
\sum_{a}F_{a}\,\e^{\,\frac{\I}{\hbar}S_{a}(E)}
\left(\e^{\,\I\frac{T_{a}}{T_{H}}\epsilon_{A}^{+}}
-\e^{\,\I\frac{T_{a}}{T_{H}}\epsilon_{C}^{+}}\right)
  \label{SemicZ} \\
&&+\sum_{a}F_{a}^*\,\e^{\,-\frac{\I}{\hbar }S_{a}(E)}
\left
(\e^{\,-\I
\frac{T_{a}}{T_{H}}\epsilon _{B}^{-}}
-\e^{\,-\I\frac{T_{a}}{T_{H}}\epsilon_{D}^{-}}\right)\Big]
\nonumber\,.
\end{eqnarray}

\section{Systems without time reversal invariance}

\subsection{Diagonal approximation}

The semiclassical representation (\ref{SemicZ}) falls into a product
over periodic orbits,
\begin{eqnarray}\label{Zprod}
Z &=&e^{\frac{i}{2}\left( \varepsilon _{A}^{+}-\varepsilon
_{B}^{-}-\varepsilon _{C}^{+}+\varepsilon _{D}^{-}\right) }Z_{0}\,,
\qquad \qquad Z_{0} =\prod_{a}z_{a}, \\
z_{a} &=&\exp \big[ \underbrace{F_{a}
\left(
\e^{\I\frac{T_{a}}{T_{H}}\epsilon_{A}^{+}}
-\e^{\,\I\frac{T_{a}}{T_{H}}\epsilon_{C}^{+}}
\right)}_{=f_{AC}^a}\e^{\,\frac{\I}{\hbar }S_{a}(E)} \nonumber\\
&& \quad
+\underbrace{\,F_{a}^*\,
\left(
\e^{\,-\I\frac{T_{a}}{T_{H}}\epsilon _{B}^{-}}
-\e^{\,-\I\frac{T_{a}}{T_{H}}\epsilon_{D}^{-}}
\right)}_{=f_{BD}^{a*}}\e^{\,-\frac{\I}{\hbar }S_{a}(E)} \big] \,.\nonumber
\end{eqnarray}
The diagonal approximation \cite{Berry} assumes that for systems
without time-reversal invariance contributions of different periodic
orbits are uncorrelated. For the generating function this means that
the energy-averaged $Z_{0}$ becomes a product of single-orbit
averages,
\begin{equation}
\langle Z_{0}\rangle_{\mathrm{diag}}=\prod_{a}\langle z_{a}\rangle\,.
\end{equation}

The energy average in $\langle z_{a}\rangle$ does away with rapid
oscillations due to the phase factors $\e^{\pm\I S(E)/\hbar}$ in
the exponent. Since $z_{a}$ is periodic in the phase
$\Phi_{a}=S_{a}(E)/\hbar $ we may average with respect to
$\Phi_a$, over a single period $2\pi$. This yields
\begin{eqnarray}
\left\langle z_{a}\right\rangle &=& \frac{1}{2\pi }\int_{0}^{2\pi
}d\Phi \exp \left(f_{AC}^{a}\e^{\I\Phi }+f_{BD}^{a*}\e^{-\I\Phi
}\right) =I_{0}\left( 2\sqrt{f_{AC}^{a}f_{BD}^{a\ast }}\right) ,
\end{eqnarray}
where $I_{0}$ is the imaginary-argument Bessel function. The
expansion $\ln I_{0}\left( y\right)
=\frac{y^{2}}{4}-\frac{y^{4}}{64}+\ldots$ gives $ \langle
Z_{0}\rangle =\exp \big[ \sum_{a}f_{AC}^{a}f_{BD}^{a*}
-\frac{1}{4}\sum_{a}( f_{AC}^{a}f_{BD}^{a*})^{2}+\ldots\big]$. In
the semiclassical limit ($T_{H}\rightarrow \infty $) it suffices
to keep only the leading quadratic term in the
exponent.\footnote{For this term the decrease of the stability
coefficients $|F_a|^2$ with the period is just compensated by the
exponential increase in the number of orbits. For all other terms
$F_a$ decreases faster; the contributions of the long orbits then
become exponentially small whereas the shortest orbits make for
corrections of the order ${\mathrm o}(T_0/T_H)$. }

 Our task is thus reduced to calculating the periodic-orbit sum
\begin{eqnarray}
&&\ln \left\langle Z_{0}\right\rangle_{\rm diag}=
\sum_{a}f_{AC}^{a}f_{BD}^{a\ast }  \label{lnZ} \\
&=&\sum_{a}|F_{a}|^{2}\left(\e^{\I\frac{T_{a}}{T_{H}}\epsilon_{A}^+}
-\e^{\I\frac{T_{a}}{T_{H}}\varepsilon _{C}^{+}}\right) \left(\e^{-\I
\frac{T_{a}}{T_{H}}\epsilon _{B}^{-}}
-\e^{-\I\frac{T_{a}}{T_{H}}\epsilon_D^-}\right)  \nonumber
\end{eqnarray}

\subsection{Diffusive vs ergodic behavior}

For ``normal'' systems the sum over periodic orbits can be done using
the well-known sum rule of Hannay and Ozorio de Almeida \cite{HOdA}
\begin{equation}
\sum_{a}|F_{a}|^{2}\,(\cdot) =\int_{T_{0}}^{\infty }
\frac{dT}{T}\,(\cdot)
\end{equation}
which expresses the approximately ergodic behavior of long orbits; short
orbits below a certain classical period $T_0$ have to be excluded.

The quasi one-dimensional character of long wires requires a
modification of that sum rule \cite{Dittrich}. The momentum
components in all directions and the transverse coordinates can
still effectively randomize after a few bounces against the
boundary, but the longitudinal coordinate takes much longer to
explore the whole length $L$ of the wire. Given chaos inducing
boundaries, each bounce will with finite probability change the
sign of the longitudinal momentum component and thus entail
diffusion along the wire.  The longitudinal coordinate becomes
randomized only for times long enough to explore the wire, i.e.,
times in excess of the diffusion (Thouless) time $T_D=L^2/D$ where
$D$ denotes the diffusion constant. Orbits with periods $T$ much
smaller than that Thouless time will have explored only the small
fraction $\sqrt{DT}/L$ of the whole length, and the inverse of
that fraction must be expected as a factor of increase of the
right hand side of the above sum rule, relative to orbits with
periods $T\gg T_D$.

Dittrich \cite{Dittrich} has determined the aforementioned factor of
increase for arbitrary values of the ratio $T/T_D$ as the integral
$\int_0^Ldx_0 p_{x_0}(x_0,T|x_0)=P(T)$ of the probability density of
return to an arbitrary point $x_0$ of departure after a time $T$, for
a one-dimensional random walk. Solution of the diffusion equation for a
wire of length $L$, led to the ``enhanced-return'' factor
\begin{equation}\label{retprob}
P(T)
= \sum_{n=0}^{\infty }\exp\Big( -\frac{\pi^{2}n^{2}T}{2T_{D}}\Big)
=\frac{1}{2}\Big[1+\vartheta_{3}\Big(0,\e^{-\pi^{2}T/2T_D}\Big)\Big]
\end{equation}
where $\vartheta _{3}(u,q) $ denotes the elliptic theta-function of
the third kind \cite{AbraSteg}. The sum rule modified for quasi one
dimensional systems thus reads
\begin{equation}
\sum_{a}|F_{a}|^{2}\,(\cdot) =\int_{T_{0}}^{\infty }
\frac{dT}{T}P(T)\,(\cdot)\, .
\end{equation}
Hannay's and Ozorio de Almeida's sum rule is restored if $T_{D}$
is so small compared with $T$ that only the $n=0$ term in
(\ref{retprob}) survives and $P(T)\sim 1$. In the opposite limit
 $T_D\gg T$ we have $P(T)\sim \sqrt{T_D/2\pi T}=L/\sqrt{2\pi DT}$ which agrees with
the above qualitative expectation.

Inasmuch as orbit periods of the order of the Heisenberg time $T_H$
determine the level statistics the time scale ratio
\begin{equation}
\zeta =\frac{\pi ^{2}T_{H}}{2T_{D}}
\end{equation}
will play an important role for the wires under study here. In
particular, the crossover from normal behavior ($\zeta\gg 1$) to
the quasi one-dimensional behavior under study here takes place
for $\zeta$ of the order unity; for $\zeta\ll 1$ the spectral
statistics must be compatible with localization.

We now invoke the modified sum rule and the identity
$\int_0^\infty \frac{dT}{T} (\e^{\I aT}-\e^{\I
bT})=\ln\frac{a}{b}$; the lower limit $T_0$ of the time integral
could be replaced by zero, accepting a negligible error ${\mathrm
o}(T_0/T_H)$. We thus get the diagonal approximation for our
generating function as
\begin{eqnarray}\label{lnZdiag}
\ln \left\langle Z\right\rangle _{\rm diag}
&\approx&\sum_{n=0}^{\infty }\int_{0}^{\infty }\frac{dT}{T}
\e^{-n^{2}\frac{\zeta T}{T_H}}\left[\e^{\I\frac{T}{T_H}(\epsilon_{A}^+
-\epsilon_B^-)}
-\e^{\I\frac{T}{T_H}(\epsilon_C^+ -\epsilon_B^-) }\right. \nonumber\\
&&\left. \hspace{2.5cm}+\,
\e^{-\I\frac{T}{T_H}(\epsilon_D^- -\epsilon_C^+)}
-\e^{-\I \frac{T}{T_H}(\epsilon_D^- -\epsilon_A^+) }\right] \nonumber\\
&=&\sum_{n=0}^{\infty }\ln \frac{\left(\I\zeta n^{2}
+\epsilon_{C}^{+}-\epsilon _{B}^{-}\right)
\left( \I\zeta n^{2}
+\epsilon_{A}^{+}-\varepsilon _{D}^{-}\right)}
{\left( \I\zeta n^{2}
+\epsilon_{A}^{+}-\epsilon_{B}^{-}\right)
\left( \I\zeta n^{2}
+\epsilon_{C}^{+}-\epsilon_{D}^{-}\right)}\,.
\end{eqnarray}
The infinite product implicitly involved here can be brought to a
closed form using
\[\prod_{n=0}^{\infty }\frac{n^{2}+a}{n^{2}+b}=\frac{\varphi \left(
a\right) }{\varphi \left( b\right) },
\quad \varphi \left( x\right)= \sqrt{x}\sinh \pi \sqrt{x}\,,
\]
whereupon we arrive at our final result for the generating function in
the diagonal approximation,
\begin{equation}
\langle Z\rangle_{\rm diag}=\e^{\frac{\I}{2}\left(
\epsilon_A^+-\epsilon_B^- -\epsilon_C^++\epsilon_D^-\right)}
\frac{\varphi \left(\frac{\epsilon_C^+ -\epsilon_B^-}{\I\zeta }\right)
\varphi \left(\frac{\epsilon_A^+ -\epsilon_D^-}{\I\zeta }\right) }
{\varphi \left(\frac{\epsilon_A^+ -\epsilon_B^-}{\I\zeta }\right)
\varphi \left(\frac{\epsilon_C^+ -\epsilon_D^-}{\I\zeta }\right) }.
\label{zuni}
\end{equation}%
In the limit $T_{D}\to 0$ or $\zeta \to \infty$ we have $\varphi(x)\to
\pi x$, and the generating function tends to its familiar form for
normal systems \cite{Oursmalltime}.

\subsection{Two-point correlator and form factor}

Substituting $\left\langle Z\right\rangle _{\mathrm{diag}}$ for $%
\left\langle Z\right\rangle $ in (\ref{CinZ}) and identifying energies
columnwise ($\parallel $) we obtain
\begin{eqnarray}\nonumber
C_{\parallel }\left(\epsilon \right)
&=&-2\sum_{n=0}^{\infty }\frac{1}{\left(
\I\zeta n^{2}+2\epsilon\right)^{2}} \\ \label{Cpar}
&=&-\frac{1}{2\epsilon^{2}}\left(\frac{1}{2}
+\frac{1}{4}\theta\cot\theta
+\frac{\I\epsilon\pi^{2}}{2\zeta\sin^{2}\theta}\right) ,
\\ \nonumber
\theta &=&\left( 1+\I\right)\pi\sqrt{\frac{\epsilon }{\zeta }}.
\end{eqnarray}
Upon taking the real part we are led to a cumbersome expression
for $R(\epsilon)$ equivalent to the earlier RMT
\cite{AndreevAltshuler} and semiclassical \cite{Dittrich}\ results
for the non-oscillatory part of the correlator. In the limit
$\zeta \rightarrow \infty $ the GUE behavior $R_{\rm
non-osc}(\epsilon) =-1/2\epsilon ^{2}$  is restored.

The crosswise ($\times $) identification of parameters, on the other
hand, entails
\begin{equation}\label{Ccross}
C_\times(\epsilon) =\frac{2\pi^{2}\e^{\I 2\epsilon }}{\epsilon\zeta
\left[ \cosh 2\pi \sqrt{\frac{\epsilon }{\zeta }}-\cos
2\pi \sqrt{\frac{\epsilon }{\zeta }}\right] }\,,
\end{equation}
and now the real part yields the oscillatory part of the
correlator $R_{\rm osc}\left(\epsilon \right)$, in agreement with
what Andreev and Altshuler \cite{AndreevAltshuler} had found
through an average over an ensemble of disordered systems. The
GUE expression $R_{\rm osc}(\epsilon) =\cos 2\epsilon /2\epsilon^{2} $
follows in the limit $\zeta \to \infty$.

For finite $\zeta$ the amplitude of oscillations of the spectral
correlation function tends to zero exponentially with $\epsilon\to
\infty$ instead of the power law characteristic of  normal
systems. That leads to qualitative changes in  the spectral form
factor $K(\tau)$ where $\tau$ is the dimensionless time, $ \;
\tau=t/T_H$. For $\tau>0$, the form factor can be defined as the
Fourier transform,
\begin{equation}\label{formfactor}
K\left( \tau \right) =\frac{1}{2\pi }\int_{-\infty +i\gamma
}^{\infty +i\gamma }e^{-i2\varepsilon \tau }C\left( \varepsilon
\right) d\varepsilon .
\end{equation}
In normal systems without time reversal invariance $K(\tau)$ experiences
discontinuity of its first
derivative at $\tau=1$ introduced by the Fourier transform of the
oscillatory part of the spectral correlation function. In quasi
one-dimensional system this discontinuity is
replaced by a smooth transition from the small-time to large-time
behavior. Another change associated with the non-oscillatory part
of the spectral correlation function is the much faster growth of
$K(\tau)$ at small $\tau$: The respective (``parallel'') part of the diagonal
form factor first deduced semiclassically in
\cite{Dittrich},
\begin{equation}\label{K<}
K_{\parallel }\left( \tau \right) =\tau \sum_{n=0}^{\infty
}e^{-\zeta n^{2}\tau }=\tau P(T_{H}\tau)\,
\end{equation}
grows like  square root rather than linearly. That faster rise
toward the saturation value unity may be seen as a (slightly
indirect) hint to localization; in the  Poissonian limit this rise
would be a jump, $K(\tau)=1$ for all $\tau>0$.

As a note of caution it has to be mentioned that, in contrast to
 normal systems without time reversal, the diagonal
approximation for the correlation functions no longer coincides
with the exact result in quasi one-dimensional systems. In
particular, the Fourier transform  of the oscillatory part
(\ref{Ccross})  is no longer zero for $\tau<1$ tending to a finite
negative value for $\tau\to+0$. Consequently the total form factor
of the diagonal approximation becomes negative for small $\tau$
although the exact form factor is known to be non-negative.

\section{Time reversal invariance}

Interesting changes arise if time-reversal invariance holds. Periodic
orbits then exist in time reversed pairs ($a,\bar{a}$) with exactly
the same action and stability coefficients. The generating function
becomes a product of contributions of different pairs, and these pairs
are uncorrelated in the diagonal approximation. Due to
$z_{a}=z_{\bar{a}}$ the generating function is squared compared to
(\ref{zuni}), the Weyl factor apart,
\begin{equation}
\langle Z\rangle_{\rm diag}=\e^{\frac{\I}{2}\left(
\epsilon _{A}^{+}-\epsilon _{B}^{-}-\epsilon _{C}^{+}+\epsilon
_{D}^{-}\right) }
\left[ \frac{\varphi \left( \frac{\epsilon
_{C}^{+}-\epsilon _{B}^{-}}{i\zeta }\right) \varphi \left( \frac{%
\epsilon _{A}^{+}-\epsilon _{D}^{-}}{i\zeta }\right) }{\varphi \left(
\frac{\epsilon _{A}^{+}-\epsilon _{B}^{-}}{i\zeta }\right) \varphi
\left( \frac{\epsilon _{C}^{+}-\epsilon _{D}^{-}}{i\zeta }\right) }%
\right] ^{2}\,. \label{zortho}
\end{equation}
 Using this generating function with
columnwise identification of arguments we find that the
non-oscillatory part of the two-point correlation function and the
small-time form factor are doubled compared to (\ref{Cpar}) and
(\ref{K<}); this is in line with Refs. \cite{AndreevAltshuler} and
\cite{Dittrich}.

Remarkably, for time-reversal invariant systems the diagonal
approximation yields no oscillatory contributions to the correlation
function, i.e., there are no terms of order $\frac{\cos 2\epsilon
}{\epsilon ^{2}}$. This can be understood as follows. In the crosswise
limit (\ref{cro}) we have
\begin{equation}
\epsilon _{C}^{+}-\epsilon _{B}^{-}=\epsilon _{A}^{+}-\epsilon
_{D}^{-}=\I 2\gamma ,\quad \gamma \to 0\;  \label{orders}
\end{equation}
 such that we can replace $\varphi(x)\to \pi x$ in the numerator of $\langle Z\rangle_{\rm diag}$;
 this gives
\begin{equation}\label{zorthocross}
  \langle Z\rangle _{{\rm diag}}\propto
(\epsilon _{C}^{+}-\epsilon _{B}^{-})^2(\epsilon _{A}^{+}-\epsilon
_{D}^{-})^2
\end{equation}
such that $\langle Z\rangle _{{\rm diag}}$  tends to zero like
$\Or(\gamma^4)$. The two derivatives w.r.t.
$\epsilon_A^+,\epsilon_B^-$ can only eliminate two factors
$\gamma$. This leaves a result that tends to zero like
$\Or(\gamma^2)$ which  means that $C_{\mathrm{diag},\times }=0$.

To derive the oscillatory component of the spectral correlator we
thus have to go beyond the diagonal approximation and take into
account correlations between the factors $z_{a}$ in the generating
function related to different periodic orbits. For the relevant
correlated orbits the differences $\left\langle
  z_{a}z_{b}\right\rangle -\left\langle z_{a}\right\rangle
\left\langle z_{b}\right\rangle ,\left\langle
  z_{a}z_{b}z_{c}\right\rangle -\left\langle z_{a}\right\rangle
\left\langle z_{b}\right\rangle \left\langle z_{c}\right\rangle $\
etc. must be non-zero. In view of the semiclassical limit this is
possible only if the respective actions have a chance to cancel,
e.g. if $S_{a}\left( E\right) \approx S_{b}\left( E\right)$ or
$S_{a}\approx S_{b}(E)+ S_{c}\left( E\right)$. Such correlations
between orbits indeed exist for chaotic dynamics; they stem from
``encounters'', i.e., places where two or more stretches of the
same orbit or of different orbits are close and almost parallel or
antiparallel to each other. By changing the connections inside
these encounters one can turn, e.g., an orbit $a$ into an orbit
$b$ with almost the same action, or split it into two orbits $b$
and $c$ whose sum of actions is close to the action of $a$. We
shall refer to such sets of correlated orbits as ``bunches''. The
simplest encounter involves two almost antiparallel orbit
stretches; the bunch it generates is the famous Sieber-Richter
pair (containing one orbit where the encounter forms a crossing in
configuration space and one where it forms an avoided crossing)
\cite{SR}. More complicated scenarios were introduced in \cite
{Oursmalltime,Ourlargetime}. It has been shown in
\cite{Ourlargetime} that taking into account both the ``diagonal''
correlations and those related to bunches gives correct
semiclassical asymptotics of the generating function and the
correlation function of normal systems. The generating function
was found as
\begin{equation}
\left\langle Z\right\rangle = \left\langle Z\right\rangle
_{\mathrm{diag} }\,\left( 1+\left\langle Z\right\rangle
_{\mathrm{off}}\right) , \label{ZTRI}
\end{equation}%
where the off-diagonal part $\langle Z\rangle_{\rm off}$ contains the
contributions of the bunches mentioned.

Let us now determine the term in $\langle Z\rangle_{\rm off}$
responsible for the leading oscillatory contribution to the
correlator. Multiplication with this term must remove one factor
$\epsilon_C^+-\epsilon_B^-\to i2\gamma$ and one factor
$\epsilon_A^+-\epsilon_D^-\to i2\gamma$ from (\ref{zorthocross}).
The product is then proportional to
$(\epsilon_A^+-\epsilon_D^-)(\epsilon_C^+-\epsilon_B^-)$ and
survives differentiation w.r.t. $\epsilon_A^+$ and $\epsilon_B^-$
and taking the limit $\gamma\to0$. For normal systems the term of
lowest order in $\frac{1}{\epsilon}$ satisfying this condition
reads (see Eqs.~(13,14) of the on-line version of
Ref.~\cite{Ourlargetime})
\begin{equation}
\left\langle Z\right\rangle _{\mathrm{off},\times
}=-\frac{4}{\left( \epsilon _{A}^{+}-\epsilon _{D}^{-}\right)
\left( \epsilon _{C}^{+}-\epsilon _{B}^{-}\right) }.  \label{ZOFF}
\end{equation}

In a quasi one-dimensional system, due to the diffusive dynamics,
Eq. (\ref{ZOFF}) would be replaced by an expression analogous to
the diagonal approximation, i.e. a summand as in
 (\ref{ZOFF}), plus terms
where the energy differences in the denominator are shifted by
finite imaginary amounts of the type $\I\zeta n^2$, with integer
$n$ as in $(\ref{lnZdiag})$. However, the shifted terms would, for
$n\neq 0$, no longer diverge  in the limit $(\times )$. Hence
combined with the above factor $\Or(\gamma^2)$ they would yield
vanishing contributions to the correlation function. Therefore the
leading term in $C_{\times }(\epsilon )$ is still due to
(\ref{ZOFF}). Substituting (\ref{ZOFF})~into (\ref{ZTRI}) and
calculating derivatives in the crosswise procedure, we obtain the
oscillatory component of the complex correlator,
\begin{equation}\label{Ccrossortho}
\lim_{\gamma \to+0}C_{\times }(\epsilon)= \frac{8\pi^{4}\,\e^{\,\I
2\epsilon }}{\epsilon^{2}\zeta^{2}\left(\cosh
2\pi\sqrt{\frac{\epsilon}{\zeta }}-\cos 2\pi \sqrt{\frac{\epsilon}
{\zeta}}\right)^{2}}\,.
\end{equation}
Its real part coincides with the RMT result for the two-point
spectral correlation function in the presence of time
reversal\cite{AndreevAltshuler}, now deduced semiclassically for
individual chaotic quasi one-dimensional systems. In the
normal-system limit $\zeta\to \infty$ (\ref{Ccrossortho}) tends to
the random-matrix expression
$\exp\left(\I2\epsilon\right)/2\epsilon^4$. On the other hand, for
all finite $\zeta$ the amplitude of oscillations diminishes
exponentially with the growth of $\epsilon$. As a consequence the
discontinuity at $\tau=1$ of the third derivative of the GOE
spectral form factor $K(\tau)$ \cite{Bible} is smoothed out in the
quasi one-dimensional case.

\ack
 Financial support of the Sonderforschungsbereich SFB/TR12 of
the Deutsche Forschungsgemeinschaft is gratefully acknowledged.

\end{document}